\documentstyle[epsf]{article}
\newcommand{\Lh}{$\Lambda$-hyperon~}
\newcommand{\pppKL}{~$pp\longrightarrow pK^+\Lambda$~}
\newcommand{\pKL}{~$pK^+\Lambda$~}
\newcommand{\Kp}{$K^+$ }

\newcommand{\e}{$ \varepsilon$}
\newcommand{\De}{$\ \Delta \varepsilon$}

\begin{document}

\title{Total Cross Section of the Reaction  \pppKL \\ 
Close to Threshold}

\author{}
\date{\today}

\maketitle
\begin{center}
 J.T.~Balewski$^{1, 4}$,
 A.~Budzanowski$^1$,
 H.~Dombrowski$^2$,
 W.~Eyrich$^8$,
 C.~Goodman$^3$,
 D.~Grzonka$^4$,
 J.~Haidenbauer$^4$,
 C.~Hanhart$^4$,
 J.~Hauffe$^8$, 
 L.~Jarczyk$^5$,
 M.~Jochmann$^6$,
 A.~Khoukaz$^2$,
 K.~Kilian$^4$,
 M.~K\"ohler$^6$,
 A.~Kozela$^1$,
 T.~Lister$^2$,
 A.~Metzger$^8$,
 P.~Moskal$^{4, 5}$,
 W.~Oelert$^4$,
 C.~Quentmeier$^2$,
 R.~Santo$^2$,
 G.~Schepers$^2$,
 U.~Seddik$^7$,
 T.~Sefzick$^4$,
 J.~Smyrski$^5$,
 M.~Soko\l owski$^5$,
 F.~Stinzing$^8$,
 A.~Strza\l kowski$^5$, 
 C.~Thomas$^2$,
 S.~Wirth$^8$,
 M.~Wolke$^4$,
 R.~Woodward$^8$,
 P.~W\"ustner$^6$,
 D.~Wyrwa$^{4, 5}$
\end{center}

\noindent
$^1$ Institute of Nuclear Physics,  Cracow, Poland \\
$^2$ IKP, Westf\"alische Wilhelms--Universit\"at, M\"unster, Germany \\
$^3$ IUCF, Bloomington, Indiana, USA \\
$^4$ IKP,  Forschungszentrum J\"ulich,  Germany \\
$^5$ Institute of Physics, Jagellonian University, Cracow, Poland \\
$^6$ ZEL,  Forschungszentrum J\"ulich,  Germany \\
$^7$ NRC, Atomic Energy Authority,  Cairo, Egypt \\
$^8$ IKP, University Erlangen-N\"urnberg,  Germany \\

 {\bf Abstract:}
\begin{flushleft}
The energy dependence of the  total cross section for the
\pppKL reaction was measured in the threshold region covering the  
excess energy range up to 7~MeV.

Existing model calculations
 describe  the slope of the measured cross sections well, 
but are too low by a factor of two to three in rate.
   
The data were used for a precise determination of
the beam momentum of the COSY-synchrotron.\\
\end{flushleft}


\section{Introduction }
The associated strangeness production in pp collisions is of fundamental 
interest and provides a possibility to study various theoretical models of the
strangeness dissociation mechanism \cite{xx}. 

In this contribution we present data on the production of the 
hyperon - kaon pair via the \pppKL elementary reaction.
In the covered excess energy range up to about 7 MeV predominantly only 
S-waves contribute to the reaction mechanism process.
   
At threshold in particular, effects of final state interactions (FSI) are 
significant and have to be taken into account.
Experimental data on the reaction \pppKL makes it possible to separate the 
effects  of  $FSI_{p-\Lambda}$ and to investigate various
meson exchange models of the nucleon-hyperon interaction ~\cite{Holz89,Maes89}.
 
The knowledge of the kaon production cross section in the elementary N-N 
interaction is important for studies of  the production of  hyper-nuclei 
in nucleon interactions with nuclei~\cite{Rudy96,Ohm97}.
 
Furthermore this elementary process is of great interest as input for 
investigations of the strangeness production mechanism in heavy ion collisions, 
which may provide information about hot, dense nuclear matter or the possible
existence of a quark-gluon plasma~\cite{Nag92}.

\section{Experiment}

\begin{figure}[tb]
\epsfysize=8.cm \centering \leavevmode \epsfverbosetrue \epsfclipon
\epsffile[9 410 515 800 ]{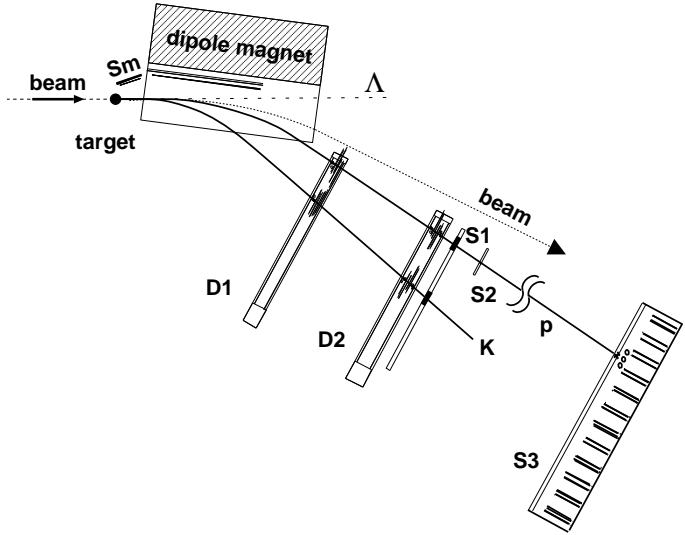}
\caption{
Sketch of the COSY-11 setup used for the \pppKL detection.
The positively charged particles, i.e. proton and kaon,
are measured by drift chambers (D1,D2)
and scintillation detectors (S1,S3). The $\Lambda$-particle and its 
decay products are not registered, only the direction of the 
$\Lambda$-particle is displayed in the figure.
The monitor scintillator (Sm) is used for the coincident detection of
elastically scattered protons.}

\label{c11_topv}
\end{figure}

The measurement of the \pppKL reaction was performed at the
  COSY-J\"ulich synchrotron~\cite{Mai94},
using the internal target facility COSY-11 \cite{Bra96}.
In fig.~\ref{c11_topv} a sketch of the experiment is shown.

\begin{figure}[h]
\epsfysize=8.cm \centering \leavevmode \epsfverbosetrue \epsfclipon
\epsffile[79 200 481 598 ]{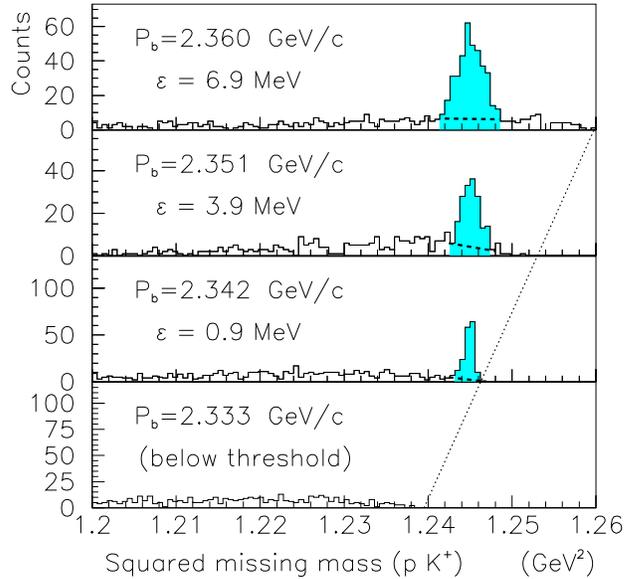}
\caption{
Squared missing mass (MM$^2$) of the pK$^+$ subsystem measured at three different nominal COSY
 beam momenta. The indicated peaks correspond to the mass of the  $\Lambda$ - particle from
 the \pppKL reaction.
The dashed line depicts the background level.
The dotted line marks the kinematical limit of the MM$^2$ spectra.}

\label{pkl_mm2}
\end{figure}

The outgoing protons and positively charged kaons were identified by means of
particle momentum reconstruction combined with a time of flight 
measurement. The  four-momentum of the unobserved \Lh was calculated from
conservation principles.
Details of the applied experimental technique are given 
elsewhere~\cite{Bra96, Bal96}.
Measurements were performed at seven excess energies between 
\e~=~0.9~MeV and \e~=~6.9~MeV according to the nominal COSY beam momentum.
Examples of (pK$^+$) missing mass  spectra, including one measurement below threshold,
are shown in fig.~\ref{pkl_mm2}. The broadening of the $\Lambda$ peak
is understood as an error propagation in the missing mass determination.
\\
\\
The geometrical acceptance of the COSY-11 detection system was 
calculated using Monte-Carlo simulations based on the code GEANT-3. The three-body 
phase space distribution was folded with the
proton-$\Lambda$ final state interaction, applying the J\"ulich model A from ref. \cite{Holz89}. 
Within the present range of the excess energy the geometrical 
acceptance
decreases with increasing beam momentum from 30\% to 5\%.
The  proton-$\Lambda$ FSI has a relative influence of less than 3\% on the acceptance and does not significantly depend on the used model.
The overall detection efficiency  ~$E_{ff}(\varepsilon)$~ is about three 
times smaller, mainly due to the decay of the \Kp mesons in flight before 
reaching the triggering  scintillator S1.

The  integrated luminosity ($I_0$)
 was determined by normalizing the simultaneously 
 recorded and extracted p+p elastic scattering to 
the cross sections measured by the EDDA collaboration~\cite{Edda97}.

\section{Determination of the excess energy} 
 
Due to the observed rapid variation of the cross section near threshold,
a precise knowledge of the beam momentum is crucial for the precision of  this experiment. 
The uncertainty in the absolute beam momentum is mainly caused by the 
uncertainty of the effective beam orbit length and amounts to
~$\Delta p/ p = 10^{-3}$. Thus, at 2.35~GeV/c beam momentum the uncertainty 
is ~$\Delta p=$~2.4~MeV/c which may cause a systematic offset in the excess energy of 
the outgoing \pKL system as large as \De~$\simeq$800~keV.
On the other hand, the small variation of the beam momentum of about 20 MeV/c 
for the present excitation function measurement between 
2.342~GeV/c and 2.360~GeV/c is controlled by measuring the revolution 
frequency in COSY with the extreme relative precision of again $10^{-3}$.
This allows a precision of 20 keV/c in momentum steps and 
the common \De~ offset is essentially the same for all seven 
measurements since these were carried out with the same beam optics.
A change of excess energy ($\varepsilon_i -\varepsilon_{i+1})$ ~between two
measurements is known with an uncertainty of only  10~keV. 

To extract the value of \De~ for the actual measurement the 
\pppKL data themselves were used.
We may write:
\begin{eqnarray}
 N /I_0 ~~~=~~~ E_{ff}(\varepsilon) \cdot \sigma(\varepsilon) 
\label{eq_n2i}
\end{eqnarray}
where $N$  stands for the number of measured \pKL events.
The left side is completely determined by the experiment, 
whereas the right hand side consists of the product of  
the known detection efficiency  and the energy dependent  cross section.
Close to threshold the cross section follows in first order the three body 
phase space distribution: 
$\sigma(\varepsilon) \sim \int \! d\rho_3 \sim \varepsilon^2 $.
Corrections due to the Coulomb interaction
 ~$f_c(q_{p_K})$~  and to the 
dominant p-$\Lambda$ final state interaction ~$f_{_{FSI}}(q_{p_\Lambda}) $~
(see Watson's model~\cite{Watson}) modify
smoothly this dependence according to:
\begin{eqnarray}
 \sigma(\varepsilon)
  ~~~\sim~~~ \int f_c(q_{p_K}) ~f_{_{FSI}}(q_{p_\Lambda})  ~d\rho_3 
\label{eq_se}
\end{eqnarray}

Only the charged particles, the proton and kaon, undergo Coulomb repulsion.
According to the approach used for the symmetric (pp) subsystem \cite{Han97}
in the ~$pp\longrightarrow pp\pi^\circ$~ reaction,  
the modification of the cross section due to the Coulomb repulsion is essentially given by 
the Coulomb penetration factor ~$f_c(q_{p_K})$~. For the
  (pK$^+$) subsystem it is given by:

\begin{eqnarray*}
\displaystyle
 f_c(q_{p_K}) = \frac{ 2\pi \gamma_q} {e^{ 2\pi \gamma_q} -1};~~~ &
\displaystyle
\gamma_q=\frac{\alpha \cdot \mu_{p_K}}{q_{p_K}};
\end{eqnarray*}
\begin{eqnarray}
\displaystyle
q_{p_K}=\sqrt{ 2 \cdot \mu_{p_K}  \cdot  \varepsilon_{p_K} };~~~~ &
\varepsilon_{p_K}=\sqrt{S_{p_K}} - m_p - m_K
\label{eq_c2g}
\end{eqnarray}
where $q_{p_K}$ is the momentum in the proton-kaon CM subsystem,
 depending on the squared sum of the p and the \Kp four-momenta  ($S_{p_K}$),
the reduced mass $\mu_{p_K}$ and the fine structure constant~$\alpha$.

In terms of the FSI in principle three interactions should be considered, namely
the subsystems: $p-\Lambda$, $K-\Lambda$, and $p-K$.
Since the first one appears to be more than an order of magnitude stronger than the 
other two \cite {Hoff,Deloff} we concentrate on 
the dominant proton-$\Lambda$ final state interaction only, which depends
on the $p-\Lambda$  momentum $q_{p\Lambda}$. 
Close to the reaction threshold $f_{_{FSI}}(q_{p_\Lambda})$ is: 
\begin{eqnarray}
\displaystyle
 f_{_{FSI}}(q_{p_\Lambda}) ~~ \sim ~~ \frac{ 1}
 { q_{p\Lambda}^2 +(  r\cdot q_{p\Lambda}^2/2 - 1/a)^2}
\label{eq_fsi}
\end{eqnarray}
with the scattering length ~a = -- 1.6~fm~ and the effective range 
parameter ~r = 2.3~fm~ 
taken from the J\"ulich model A given in ref.~\cite{Holz89}.

\begin{figure}[h]
\epsfysize=4.cm \centering \leavevmode \epsfverbosetrue \epsfclipon
\epsffile[28 494 280 659 ]{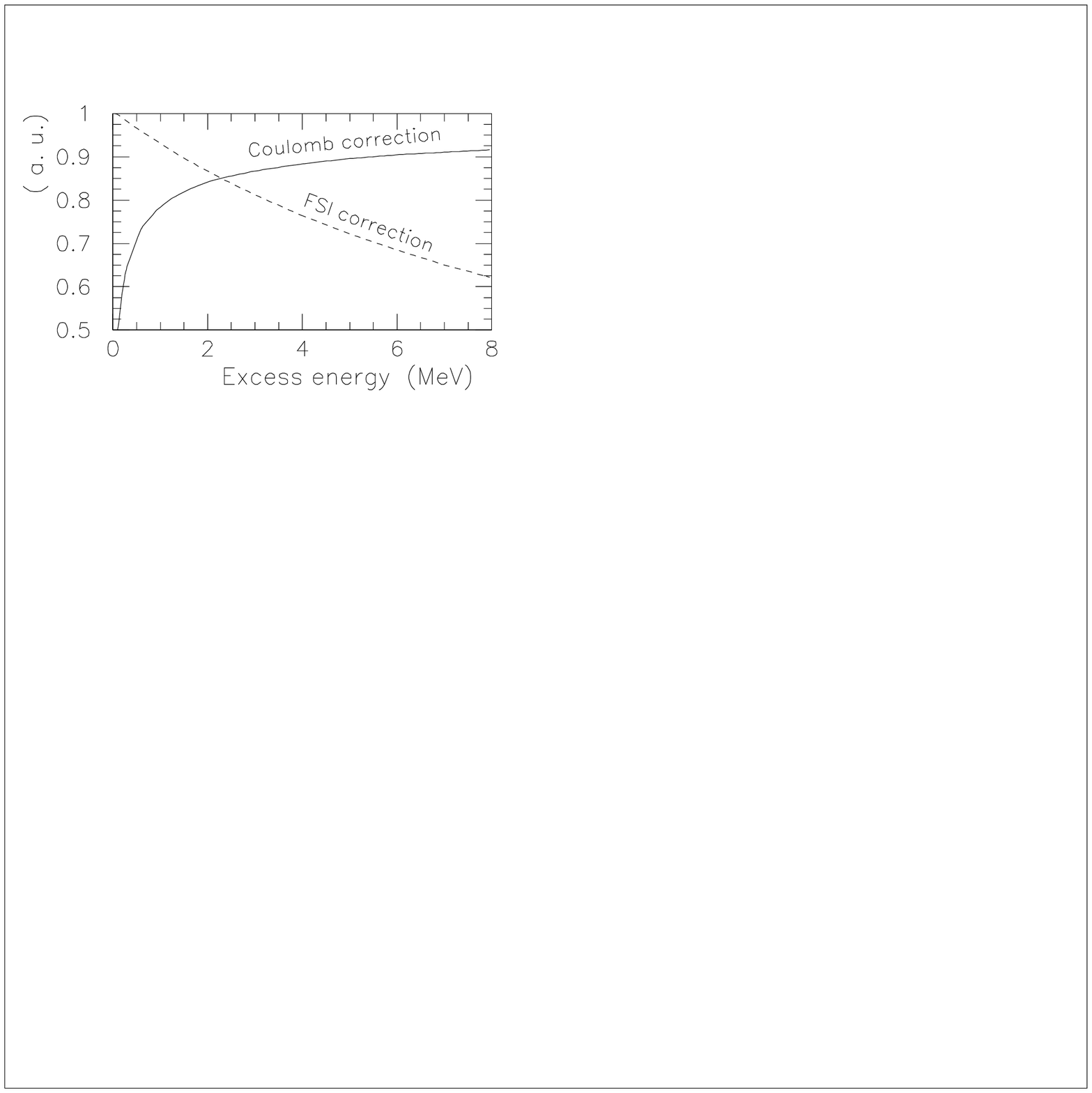}
\caption{ 
Shape distribution of the Coulomb and FSI corrections  in arbitrary units, 
calculated as 
 $ { \int \! f_c\, d\rho_3}/ {\int \! d\rho_3}$ ~~and~~ 
 ${ \int \!f_{_{FSI}} d\rho_3} /{\int\! d\rho_3}$
 resulting from 
equations~(\ref{eq_c2g}) and (\ref{eq_fsi}), respectively.}
  \label{coul+fsi+eff}
\end{figure}

Qualitatively the influence  of the two correction functions
on the energy dependence of the total cross section,
shown in fig.~\ref{coul+fsi+eff}\ , are arbitrarily normalized.\\

Denoting the nominal value of the
excess energy  by $\tilde{\varepsilon}$, calculated from the beam momentum given to 
us by the COSY team, we rewrite equation \ref{eq_n2i} to :
\begin{eqnarray}
 N /I_0 ~~=~~   E_{ff}(\tilde{\varepsilon}-\Delta \varepsilon)
 \cdot \sigma(\tilde{\varepsilon}-\Delta \varepsilon) 
 \label{eq_re}
\end{eqnarray}
 Figure~\ref{pkl_n2i} compares the threshold behavior with and without 
both Coulomb distortion effects and the final state interaction. It is obvious that
the shape of the data is much better described when including these effects.
The offset in the excess energy  was obtained from a fit of equation~(\ref{eq_re}) 
[inserting equations \ref{eq_se}--\ref{eq_fsi} ] to the data 
resulting in a smaller real excess energy by ~\De=220~$\pm$~60~keV~ with respect to 
the one calculated from the nominal beam momentum.
The equivalent shift in the nominal COSY beam momentum is 
 $\Delta$p=660$\pm$180~keV/c and is still four times smaller than the  $\Delta p/$p uncertainty typically estimated for the 
COSY beam.

\newpage
\section{Results and Discussion}

The obtained values for the \pppKL cross sections at the corrected  
excess energies  $(\tilde{\varepsilon}_i-\Delta \varepsilon)$
and with  the FSI taken into account for the
acceptance determination, are shown in table~1.
The statistical errors range from  5\%  to  9\%  for all data points.  
The systematic error includes a 4\% error in absolute normalization of the
EDDA ~$pp\longrightarrow pp$~data~\cite{Edda97} and
the uncertainties of the COSY-11~ acceptance determination for both
  ~~$pp\longrightarrow pp$~~  and  ~~\pppKL~~ reactions; 
each being estimated to be 5\%.

\begin{figure}[htb]
\epsfysize=9.cm \centering \leavevmode \epsfverbosetrue \epsfclipon
\epsffile[33 401 283 658 ]{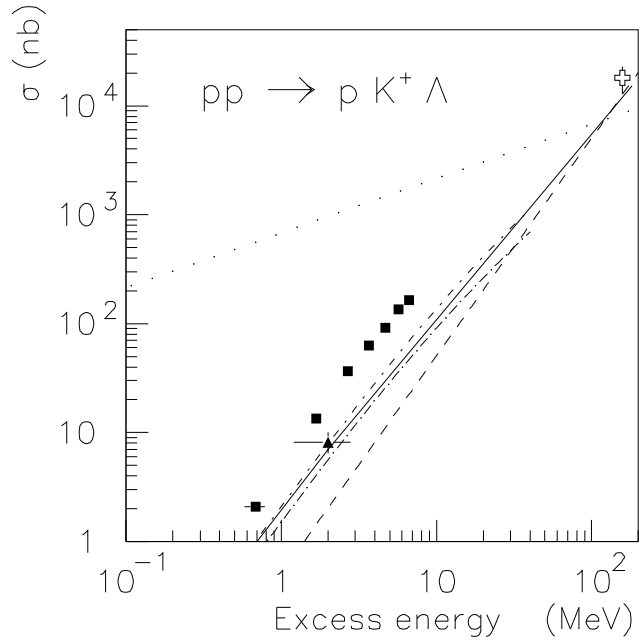}
\caption{Data and different models describing the energy dependence 
of the \pppKL total
cross section. 
The parametrization of Randrup et al. is depicted as a dotted line,
of Sch\"urman et al. as a dashed one, of Sibirtsev et al. as a solid and
of F\"aldt and Wilkin as two types of dash-dotted. Tha data of Fickinger are marked as a cross, of Balewski at al. as a triangle and the present one as squares. } 
 \label{pkl_cros_mod}
\end{figure}

Different models for the energy dependence of the \pppKL cross section
have been suggested. As can be seen from fig.~\ref{pkl_cros_mod}, where the present
and previously published data~\cite{Bal96,Fic62} are depicted in two dimensional
logarithmic scales, both the square root  and the quadratic excess energy approximations 
of J. Randrup and C.M. Ko \cite{RaKo} and B. Sch\"urmann and W. Zwermann \cite{Zwe88}, 
respectively, fail by more than an order of magnitude in describing the present data 
at threshold.

The data point from our previous measurement~\cite{Bal96} appears either to be too 
low in cross section by a factor of about 2.5 and/or is associated to a wrong excess 
energy. Utilizing fully the uncertainty of the COSY beam momentum as
~$\Delta p/ p = 10^{-3}$, which converts to an uncertainty of the excess energy to
~$\Delta \varepsilon$ = $\pm$~800~keV, the former data point is still consistent with the present
results. With only one measurement the determination of the real beam momentum 
was not possible. In addition, a 30\% increase for the present value of the cross section is due to an improved
determination of the defocusing features of the fringe field of the dipole magnet.
\\
\\
 Recently, G. F\"aldt and C. Wilkin~\cite{FeWi96} presented a one pion exchange model 
which assumes a dominant role of the second  $S_{11}$ resonance, the N$^*$ (1650),
for the \Kp production in p+p interaction.
This interpretation is analogous to the $\eta$ production,
which is supposed to be mediated  
mainly via the first $S_{11}$ resonance, the N$^*$ (1535).
These two N$^*$ resonances are distinguished by a rather large branching ratio
into K~$\Lambda$ and $\eta$~N, respectively. If the two systems $\eta$~N and
K~$\Lambda$ are dominated by these two $N^*$ resonances, the forms of the production
operators and all the spin - angular momentum algebra are identical, and the 
observation of an $\eta$ or a K merely tags which of the two $S_{11}$ resonances 
has been excited~\cite{FeWi96}. The two curves in fig.~\ref{pkl_cros_mod} are due 
to different $\Lambda$p scattering parameters used by 
G. F\"aldt and C. Wilkin ~\cite{FeWi96}, within a common factor of two to three the
data are reproduced. 

A similar, more extended model was presented by Sibirtsev~\cite{Sib96}.
In his calculation of the ~$NN\longrightarrow$ NYK cross section the one-pion
as well as the one-kaon exchange diagrams are included. The amplitudes of the
elementary processes ($\pi N \longrightarrow$ KY, KN $\longrightarrow$ KN) are
based on phenomenological parametrizations. Contrary to the model of
G. F\"aldt and C. Wilkin~\cite{FeWi96}, effects of the FSI are, however, not 
taken into account in Sibirtsev's calculations~\cite{Sib96}.
Beside the $S_{11}$ $N^*$(1650) these authors also 
include other $N^*$ resonances with masses lower than 2000 MeV
which decay into kaon - hyperon channels; but they neglect interference terms.   
Again the absolute rate differs by a factor of 2.5 whereas the
shape of the experimental excitation function is well reproduced.
\\

\begin{figure}[htb]
\epsfysize=7.cm \centering \leavevmode \epsfverbosetrue \epsfclipon
\epsffile[21 416 289 652 ]{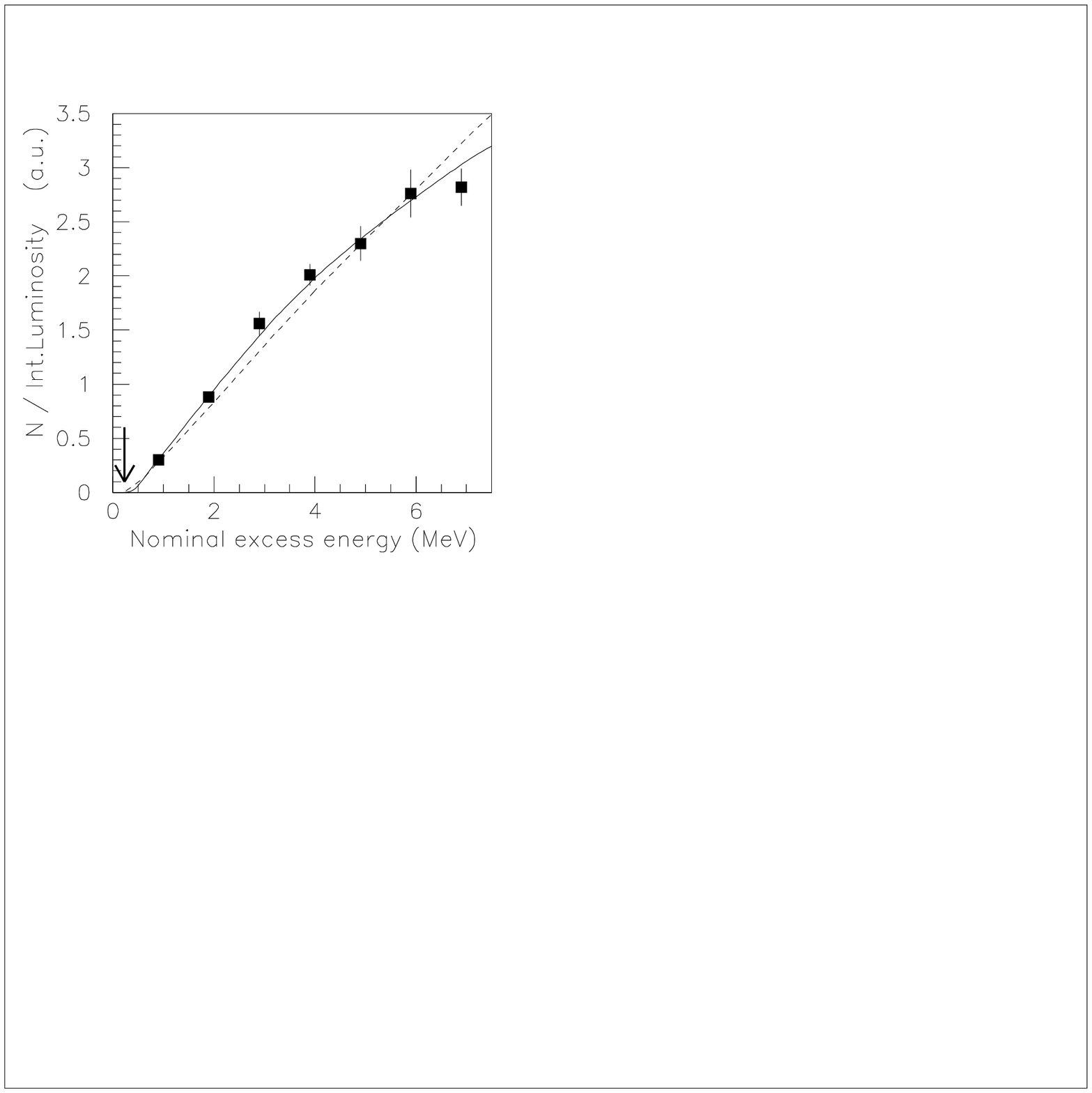}
\caption{
Determination of the beam momentum by means of the
experimentally measured  \pppKL event rate.
The solid line depicts the best fit of ~\De~ from equation \ref{eq_re} to the data.
The dashed one is the pure phase space prediction.
The arrow corresponds to the resulting offset of ~\De$=220 \pm 60~keV$.
 }
\label{pkl_n2i}
\end{figure}

In conclusion, total \pppKL cross sections have been measured in the threshold region
for excitation energies less than 7 MeV.
It has been shown that 
 the slope of the measured values  for the absolute cross section
 follows rather
the prediction including the p-$\Lambda$ FSI than the pure phase space
 calculation, see fig.~\ref{pkl_n2i} for comparison.
Model calculations including light and
heavy boson exchange as well as intermediate $N^*$ resonances seem to give a good account 
for the shape of the excitation function, but differ in magnitude by a factor of
two to three.
Further investigations on both the experimental side and the theoretical
description are required for 
a deeper understanding of the reaction mechanism and the strangeness dissociation
processes. 
For this measurements the average value of the COSY beam
momentum was determined such that a precision of 60 keV for the excess energy
is achieved.
This procedure should be applied in each close to threshold measurement.
\\
\\
\\
\begin{flushleft}
\textbf{Acknowledgements:}
\\
We appreciate the work provided by the COSY operation team for the good 
cooperation and for delivering the excellent proton beam.

We would like to thank Prof. Dr. C. Wilkin for inspiring and very helpful
discussions.

This work is based on parts of the Doctoral Thesis of G. Schepers.
 
The research project was supported by the BMBF, the Polish Committee for 
Scientific Research, and the Bilateral Cooperation between Germany and Poland
represented by the Internationales B\"uro DLR for the BMBF.
\end{flushleft} 
\newpage

\begin{table}[htb]
\begin{minipage}{16cm}
\begin{center}

\begin{tabular}{|c||c|c||c c c||c|}
  \hline
\e~~\footnote{known with an accuracy of 60~keV})
 &  \multicolumn{2}{|c||}{   events}  &  \multicolumn{3}{|c|| }{$\sigma$}  &  {$\sigma/\varepsilon^2$} \\
(MeV) & \pKL & background &   \multicolumn{3}{|c|| }{(nb)}                  &   \ (nb/MeV$^2$)             \\
  \hline
\hline
 0.68 &          216 &           27 &   2.1 & $\pm$  0.2  & & 4.54   \\
 1.68 &          598 &           58 &  13.4 & $\pm$  0.7  & & 4.75   \\
 2.68 &          378 &           58 &  36.6 & $\pm$  2.6  &  & 5.10   \\
 3.68 &          836 &          151 &  63.0 & $\pm$  3.1  & $\pm$ 14\%  & 4.65   \\
 4.68 &          412 &           68 &  92.2 & $\pm$  6.5  &  & 4.21   \\
 5.68 &          290 &           39 & 135. & $\pm$ 11.   &  & 4.18   \\
 6.68 &          449 &           76 & 164. & $\pm$  10.   &  & 3.68   \\

\hline
\end{tabular}
\end{center}
\end{minipage}
\caption{Total cross section for the \pppKL reaction.
For each excess energy (\e) the number of the \pKL events,
the estimated number of background events and the extracted  cross section are given.
The statistical and systematical errors are listed, respectively. 
The last column shows the ratio of the cross section over the square of excess energy, 
the constant value of \mbox{4.4 $\pm$ 0.7} demonstrates the approximate
phase space like increase of the total cross section with increasing excess energy.}
\end{table}


\end{document}